\documentclass[a4paper,11pt]{article}
\usepackage{pos}

\title{Rare K decays off and on the lattice}

\author[a]{Stefan Schacht}
\author*[b]{Amarjit Soni}

\affiliation[a]{Department of Physics and Astronomy, University of Manchester, Manchester M13 9PL, United Kingdom}

\affiliation[b]{Physics Department, Brookhaven National Laboratory, Upton, NY 11973, USA}

\emailAdd{stefan.schacht@manchester.ac.uk}
\emailAdd{adlersoni@gmail.com}

\abstract{The importance of rare $K$ decays especially in the context of a kaon unitarity triangle (KUT) is emphasized. The decay $K_L \to \pi^0 \nu \bar \nu$ is theoretically very clean but experimentally extremely challenging. The Standard Model prediction $\mathcal{B}\sim 3 \times 10^{-11}$ is still about two orders of magnitude away from the current experimental upper bound. One way to continue to make progress  towards the construction of a KUT is by improving the accuracy in the calculation of $\varepsilon'$. Another  way which is the primary focus here is via studies of $K^0 \to \pi^0 \mu^+ \mu^-$.  LHCb, J-PARC, the proposed HIKE project, phenomenology, and in fact precision studies on the lattice can all play a very important role in this context.}

\FullConference{The 40th International Symposium on Lattice Field Theory (Lattice 2023)\\
July 31st - August 4th, 2023\\
Fermi National Accelerator Laboratory\\}

\begin{document}

\maketitle

\section{Introduction and Motivation}

The experiment of Christenson, Cronin, Fitch and Turlay~\cite{Christenson:1964fg} showed for the first time that $K_L$ does decay to a two-pion final state. Thereby the experiment gave the first evidence for CP violation. It is actually indirect CP  violation since the mixing between $K^0$ and $\bar{K}^0$ is primarily responsible for the seen effect. Nevertheless, its observation has the profound consequence that CP is not a symmetry of nature. Naturalness arguments then suggest that new physics is likely to entail new CP-odd phases. Searching for such phases is perhaps one of the most promising ways to probe for new physics.

Our primary focus here is to probe for new physics by the construction of the kaon unitarity triangle (KUT)~\cite{Lehner:2015jga} and below we give a brief review of the ongoing and future efforts towards constructing it with rare kaon decays. The current construction of the unitarity triangle is dominated by $B$ physics, and a construction only from kaon decays would represent a crucial intergenerational test of our understanding of flavor physics and thereby the Standard Model~(SM). Any deviation between the triangle constructions with $K$ and $B$ decays, respectively, would be a clear sign for new physics.
The opportunities at future kaon facilities obtained a lot of recent attention~\cite{Anzivino:2023bhp}, and include the potential to discover 
a light dark sector~\cite{Goudzovski:2022vbt}.
A detailed summary of SM predictions can be found in Ref.~\cite{Cirigliano:2011ny}.

The rare decay $K_L \to \pi^0 \nu \bar \nu$ is an important target for the construction of the KUT and other purposes. It violates CP~\cite{Littenberg:1989ix} and is theoretically very clean; it is dominated by the top quark and therefore it is a safe probe for short distance physics. However, its measurement is unfortunately experimentally exceedingly challenging. After many years and a lot of progress, the KOTO experiment at J-PARC has obtained an impressive  upper bound of $3.0\cdot 10^{-9}$~\cite{KOTO:2018dsc}. This is still about two orders of magnitude above the SM~\cite{Buras:2015qea},~and a lot more work is anticipated~\cite{Aoki:2021cqa}. 

The challenges of measuring $K_L \to \pi^0 \nu \bar \nu$ are why it would be extremely useful to improve constraints also on the related modes involving lepton pairs: $K^0 \to \pi^0 l^+ l^-$.  These are theoretically less \lq\lq{}clean\rq\rq{}, so theoretical input becomes rather important but their experimental detection can be much less challenging.

Below, in sections~\ref{sec:Kpluspinunubar}--\ref{sec:Kmumu} we give an overview over current developments in key rare decay channels including the above. We conclude in Sec.~\ref{sec:conclusions}.

\section{$K^+ \to \pi^+ \nu \bar \nu$ \label{sec:Kpluspinunubar}}

This rare decay of the charged kaon is an important pillar of the KUT~\cite{Lehner:2015jga}. While a large fraction of the amplitude is dominated by short-distance physics, in particular the top-quark contribution, giving sensitivity to ${V^*_{ts}V_{td}}$, unfortunately it does also receive some 
long-distance contributions from charm- and up-quark intermediate states. These have been calculated in chiral perturbation theory~\cite{Isidori:2005xm} and are subject to on-going lattice~\cite{Christ:2019dxu} and phenomenological~\cite{LunghiSoni:InPreparation} studies. 
BNL experiments~\cite{BNL-E949:2009dza} saw a few candidate events and obtained a result, 
\begin{align}
\mathcal{B}(K^+ \to \pi^+ \nu \bar \nu)^{\mathrm{BNL}} &= (17.3^{+11.5}_{-10.5})\times 10^{-11}\,,
\end{align}
which is $\sim 1 \sigma$ above the SM prediction~\cite{Buras:2015qea}
\begin{align}
\mathcal{B}(K^+ \to \pi^+ \nu \bar \nu)^{\mathrm{SM}} &= (8.4\pm 1.0)\times 10^{-11}\,.
\end{align}
The experimental effort continued at CERN with the NA62 experiment, and with several years of improvements and progress, NA62 observed $K^+\rightarrow \pi^+\nu\bar\nu$ with a significance of $3.4\sigma$, resulting in the branching ratio measurement~\cite{NA62:2021zjw}
\begin{align}
\mathcal{B}(K^+\rightarrow \pi^+ \nu \bar{\nu})^{\mathrm{NA62}} = \left( 10.6^{+4.0}_{-3.4}\pm 0.9\right)\cdot 10^{-11}\,, 
\end{align}
in agreement with the SM.
Fig.~4 in Ref.~\cite{Ceccucci:2021gpl} summarizes the history of theoretical predictions and experimental measurements of $\mathcal{B}(K^+\rightarrow \pi^+\nu\bar{\nu})$. It is interesting to observe that improvements of the bound on $\mathcal{B}(K^+\rightarrow \pi^+ \nu \bar{\nu})$ from around $10^{-9}$ to around  $10^{-11}$ took $\sim$20 years. This underscores the
importance of continuous efforts on long time scales needed in kaon physics.

\section{The gold-plated decay $K_L \to \pi^0 \nu \bar \nu$ \label{sec:K0pinunubar}}

The decay $K_L \to \pi^0 \nu \bar \nu$ violates CP~\cite{Littenberg:1989ix}.
In the SM, this is a consequence of all three generations of up-type quarks $u$, $c$, and $t$ entering the electroweak loop inducing the $s\rightarrow d$ FCNC transition.
However, the top quark contribution dominates completely  and therefore the process is clearly dominated by short-distance physics.
Thus, a measurement of its rate would cleanly give the Kobayashi-Maskawa phase necessary for CP violation to occur. However, from the experimental point of view, $K_L\rightarrow \pi^0\nu\bar{\nu}$ is a very challenging process. After many years of progress, the KOTO experiment at J-PARC has achieved a ($90\%$ CL)  upper bound of~\cite{KOTO:2018dsc}
\begin{align}
\mathcal{B}(K_L \to \pi^0 \nu \bar \nu)^{\mathrm{KOTO}} \le (3.0 \times 10^{-9})\,. 
\end{align}
This upper bound is still some two orders of magnitude larger than the SM prediction~\cite{Buras:2015qea} 
\begin{align}
\mathcal{B}(K_L \to \pi^0 \nu \bar \nu)^{\mathrm{SM}} = (3.4 \pm 0.6) \times 10^{-11}\,,
\end{align}
and consequently it could take some time before we reach the discovery of this mode.
It is important therefore to tighten the constraints on this process as much as possible and 
a lot more experimental work is planned in the future~\cite{Aoki:2021cqa}.

\section{$K^0 \to \pi^0 e^+ e^-$ \label{sec:Kpiee}}

This charged counterpart of the gold-plated mode discussed in the last section  has been of interest for a very  long time~\cite{Donoghue:1987awa,Ecker:1987qi, Buchalla:2003sj,AtwoodSoni:InPreparation, DAmbrosio:1998gur, DAmbrosio:2018ytt}.
For the case of a $\pi^0 e^+ e^-$ final state, the experimental detection faces a daunting challenge by the so-called \lq\lq{}Greenlee background\rq\rq{}~\cite{Greenlee:1990qy}. The problem in the actual detection of the $\pi^0$ is that~\cite{ParticleDataGroup:2022pth}
\begin{align}
\mathcal{B}(K_L \to e^+ e^- \gamma \gamma) \approx (5.95\pm 0.33) \cdot 10^{-7}\,,
\end{align}
which is some four orders of magnitude larger than the SM prediction~\cite{Buchalla:2003sj}
\begin{align}
\mathcal{B}(K_L \to \pi^0 e^+ e^-) = \left(3.2^{+1.2}_{-0.8}\right)\cdot 10^{-11}\,.
\end{align}
Moreover, even if one could detect $\pi^0 e^+ e^-$ then the portion of the amplitude that comes from one virtual photon (or $Z$) going to $e^+e^-$ is CP violating but if the pair comes from two virtual photons the contribution is CP-conserving. So a good theoretical understanding becomes mandatory. And that is where modern lattice methods become very relevant along with, of course, phenomenology.

\section{$K^0\rightarrow \pi^0\mu^+\mu^-$ \label{sec:Kpimumu}}

This mode is very interesting as LHCb has a very large flux of $K_S$ which they can use to study decays to final states like $\mu^+ \mu^-$ and $\pi^0 \mu^+ \mu^-$~\cite{AlvesJunior:2018ldo}. $K_L$ could be studied, for example, at J-PARC and at the proposed HIKE experiment at CERN~\cite{HIKE:2022qra, HIKE:2023ext}.  In this case, the Greenlee background is three (rather than four) orders of magnitude larger than the potential signal for $K_L$ decays, as~\cite{ParticleDataGroup:2022pth}
\begin{align}
\mathcal{B}(K_L \to \mu^+ \mu^- \gamma \gamma) &= \left(1.0^{+0.8}_{-0.6}\right)\cdot 10^{-8} \,,
\end{align}
compared to the SM prediction~\cite{Isidori:2004rb}
\begin{align}
\mathcal{B}(K_L \to \pi^0 \mu^+ \mu^-)^{\mathrm{SM}} = \left(1.5\pm 0.3\right)\cdot 10^{-11} \,.
\end{align}
On the phenomenological front considerable work has already been done~\cite{Isidori:2004rb,DAmbrosio:2022kvb} and more is in progress~\cite{SchachtSoni:InPreparation}. Prospects for precise lattice studies are quite good given that the RBC-UKQCD collaboration has and is doing several related studies~\cite{Christ:2015aha, Christ:2019dxu}. In particular, this means that lattice methods may be used to calculate the CP violating and CP conserving contributions to $\mu^+ \mu^-$ or $e^+ e^-$  from one photon and two photons.

From the results in Ref.~\cite{Isidori:2004rb} one obtains the ratio of the CP-conserving contribution over the total branching ratio as 
\begin{align}
\mathcal{B}(K_L\rightarrow \pi^0\mu^+\mu^-)^{\mathrm{CPC}}/\mathcal{B}(K_L\rightarrow\pi^0 
\mu^+ \mu^-) \sim 0.3\,.
\end{align}
In order to distinguish CP-conserving and CP-violating contributions to $K_L\rightarrow \pi^0l^+l^-$
experimentally, one can employ the Dalitz distribution and asymmetry observables~\cite{Donoghue:1987awa, Heiliger:1992uh, Donoghue:1994yt, Diwan:2001sg, Buchalla:2003sj}. 

It is known that the decays $K_L\rightarrow \pi^0 e^+e^-$ and $K_L\rightarrow \pi^0\mu^+\mu^-$ are complementary regarding their constraints on new physics models, as together they can be used to distinguish between (pseudo)scalar and (axial)vector new physics operators~\cite{Mescia:2006jd}.
Furthermore, they allow the test of lepton flavor universality in kaon decays~\cite{Crivellin:2016vjc, DAmbrosio:2022kvb}.
The effect of future measurements of $\mathcal{B}(K_L\rightarrow \pi^0l^+l^-)$ on the constraints of Wilson coefficients of new physics models are shown in Ref.~\cite{DAmbrosio:2023irq}.

\section{$K\rightarrow \mu^+\mu^-$ \label{sec:Kmumu}}

Recently, it has been shown that through time-dependent measurements of $K\rightarrow \mu^+\mu^-$ it would in principle be possible to separate  short- and long-distance physics, thereby enabling a new technique for the determination of the unitarity triangle height~$\eta$~\cite{DAmbrosio:2017klp, Dery:2021mct, Brod:2022khx, Dery:2022yqc}, see for a short summary Ref.~\cite{Schacht:2023yav}. Experimentally, this method is very challenging, see for first ideas Ref.~\cite{Marchevski:2023kab}.
However, already time-integrated measurements of $\mathcal{B}(K_S\rightarrow \mu^+\mu^-)$~\cite{LHCb:2020ycd} are constraining new physics models~\cite{Dery:2021vql}.
The prospects for future lattice results are promising~\cite{Christ:2022rho, Christ:2020bzb}.

\section{Conclusions  \label{sec:conclusions}}

In the upcoming era of precision flavor physics, it will be  important to 
construct the unitarity triangle primarily  from kaon decays only, in order to further test our
understanding of CP violation in the SM. This has special importance also due to the on-going tension in $V_{cb}$ and $V_{ub}$ which make a unitarity triangle construction independent of those quantities desirable, see also Ref.~\cite{Buras:2021nns}.
Any deviations from the current construction of the unitarity triangle, which is dominated by $B$ physics, will be a strong sign of new physics.
We reiterate that the gold-plated channel $K_L\rightarrow \pi^0\nu\bar{\nu}$ can be very powerfully constrained by the semileptonic decays $K^0 \to \pi^0 l^+l^-$ through their shared dependence on the height $\eta$ of the unitarity triangle, along with $K^+\rightarrow \pi^+\nu\bar{\nu}$ through the Grossman-Nir bound~\cite{Grossman:1997sk}.

\section*{Acknowledgments}
S.S. is supported by a Stephen Hawking Fellowship from UKRI under reference EP/T01623X/1 and the STFC research grants ST/T001038/1 and ST/X00077X/1. 

\bibliographystyle{JHEP}
\bibliography{draft.bib}

\providecommand{\href}[2]{#2}\begingroup\raggedright\begin{thebibliography}{10}

\bibitem{Christenson:1964fg}
J.H.~Christenson, J.W.~Cronin, V.L.~Fitch and R.~Turlay, \emph{{Evidence for
  the $2\pi$ Decay of the $K_2^0$ Meson}},
  \href{https://doi.org/10.1103/PhysRevLett.13.138}{\emph{Phys. Rev. Lett.}
  {\bfseries 13} (1964) 138}.

\bibitem{Lehner:2015jga}
C.~Lehner, E.~Lunghi and A.~Soni, \emph{{Emerging lattice approach to the
  K-Unitarity Triangle}},
  \href{https://doi.org/10.1016/j.physletb.2016.04.064}{\emph{Phys. Lett. B}
  {\bfseries 759} (2016) 82}
  [\href{https://arxiv.org/abs/1508.01801}{{\ttfamily 1508.01801}}].

\bibitem{Anzivino:2023bhp}
G.~Anzivino et~al., \emph{{Workshop summary -- Kaons@CERN 2023}},  in
  \emph{{Kaons@CERN 2023}}, 11, 2023
  [\href{https://arxiv.org/abs/2311.02923}{{\ttfamily 2311.02923}}].

\bibitem{Goudzovski:2022vbt}
E.~Goudzovski et~al., \emph{{New physics searches at kaon and hyperon
  factories}}, \href{https://doi.org/10.1088/1361-6633/ac9cee}{\emph{Rept.
  Prog. Phys.} {\bfseries 86} (2023) 016201}
  [\href{https://arxiv.org/abs/2201.07805}{{\ttfamily 2201.07805}}].

\bibitem{Cirigliano:2011ny}
V.~Cirigliano, G.~Ecker, H.~Neufeld, A.~Pich and J.~Portoles, \emph{{Kaon
  Decays in the Standard Model}},
  \href{https://doi.org/10.1103/RevModPhys.84.399}{\emph{Rev. Mod. Phys.}
  {\bfseries 84} (2012) 399} [\href{https://arxiv.org/abs/1107.6001}{{\ttfamily
  1107.6001}}].

\bibitem{Littenberg:1989ix}
L.S.~Littenberg, \emph{{The CP Violating Dacay K0(L) ---\ensuremath{>} pi0
  Neutrino anti-neutrino}},
  \href{https://doi.org/10.1103/PhysRevD.39.3322}{\emph{Phys. Rev. D}
  {\bfseries 39} (1989) 3322}.

\bibitem{KOTO:2018dsc}
{\scshape KOTO} collaboration, \emph{{Search for the $K_L \!\to\! \pi^0 \nu
  \overline{\nu}$ and $K_L \!\to\! \pi^0 X^0$ decays at the J-PARC KOTO
  experiment}},
  \href{https://doi.org/10.1103/PhysRevLett.122.021802}{\emph{Phys. Rev. Lett.}
  {\bfseries 122} (2019) 021802}
  [\href{https://arxiv.org/abs/1810.09655}{{\ttfamily 1810.09655}}].

\bibitem{Buras:2015qea}
A.J.~Buras, D.~Buttazzo, J.~Girrbach-Noe and R.~Knegjens, \emph{{$ {K}^{+}\to
  {\pi}^{+}\nu \overline{\nu} $ and $ {K}_L\to {\pi}^0\nu \overline{\nu} $ in
  the Standard Model: status and perspectives}},
  \href{https://doi.org/10.1007/JHEP11(2015)033}{\emph{JHEP} {\bfseries 11}
  (2015) 033} [\href{https://arxiv.org/abs/1503.02693}{{\ttfamily
  1503.02693}}].

\bibitem{Aoki:2021cqa}
K.~Aoki et~al., \emph{{Extension of the J-PARC Hadron Experimental Facility:
  Third White Paper}},  \href{https://arxiv.org/abs/2110.04462}{{\ttfamily
  2110.04462}}.

\bibitem{Isidori:2005xm}
G.~Isidori, F.~Mescia and C.~Smith, \emph{{Light-quark loops in K
  ---\ensuremath{>} pi nu anti-nu}},
  \href{https://doi.org/10.1016/j.nuclphysb.2005.04.008}{\emph{Nucl. Phys. B}
  {\bfseries 718} (2005) 319}
  [\href{https://arxiv.org/abs/hep-ph/0503107}{{\ttfamily hep-ph/0503107}}].

\bibitem{Christ:2019dxu}
{\scshape RBC, UKQCD} collaboration, \emph{{Lattice QCD study of the rare kaon
  decay $K^+\to\pi^+\nu\bar{\nu}$ at a near-physical pion mass}},
  \href{https://doi.org/10.1103/PhysRevD.100.114506}{\emph{Phys. Rev. D}
  {\bfseries 100} (2019) 114506}
  [\href{https://arxiv.org/abs/1910.10644}{{\ttfamily 1910.10644}}].

\bibitem{LunghiSoni:InPreparation}
E.~Lunghi and A.~Soni, \emph{{in preparation}}, .

\bibitem{BNL-E949:2009dza}
{\scshape BNL-E949} collaboration, \emph{{Study of the decay $K^+\to\pi^+\nu
  \bar\nu$ in the momentum region $140 < P_\pi < 199$ MeV/c}},
  \href{https://doi.org/10.1103/PhysRevD.79.092004}{\emph{Phys. Rev. D}
  {\bfseries 79} (2009) 092004}
  [\href{https://arxiv.org/abs/0903.0030}{{\ttfamily 0903.0030}}].

\bibitem{NA62:2021zjw}
{\scshape NA62} collaboration, \emph{{Measurement of the very rare
  $K^+\rightarrow \pi^+ \nu\bar{\nu}$ decay}},
  \href{https://doi.org/10.1007/JHEP06(2021)093}{\emph{JHEP} {\bfseries 06}
  (2021) 093} [\href{https://arxiv.org/abs/2103.15389}{{\ttfamily
  2103.15389}}].

\bibitem{Ceccucci:2021gpl}
A.~Ceccucci, \emph{{Rare Kaon Decays}},
  \href{https://doi.org/10.1146/annurev-nucl-102419-054905}{\emph{Ann. Rev.
  Nucl. Part. Sci.} {\bfseries 71} (2021) 113}.

\bibitem{Donoghue:1987awa}
J.F.~Donoghue, B.R.~Holstein and G.~Valencia, \emph{{K(L) ---\ensuremath{>} pi0
  e+ e- as a Probe of CP Violation}},
  \href{https://doi.org/10.1103/PhysRevD.35.2769}{\emph{Phys. Rev. D}
  {\bfseries 35} (1987) 2769}.

\bibitem{Ecker:1987qi}
G.~Ecker, A.~Pich and E.~de~Rafael, \emph{{K ---\ensuremath{>} pi Lepton+
  Lepton- Decays in the Effective Chiral Lagrangian of the Standard Model}},
  \href{https://doi.org/10.1016/0550-3213(87)90491-3}{\emph{Nucl. Phys. B}
  {\bfseries 291} (1987) 692}.

\bibitem{Buchalla:2003sj}
G.~Buchalla, G.~D'Ambrosio and G.~Isidori, \emph{{Extracting short distance
  physics from K(L,S) ---\ensuremath{>} pi0 e+ e- decays}},
  \href{https://doi.org/10.1016/j.nuclphysb.2003.09.010}{\emph{Nucl. Phys. B}
  {\bfseries 672} (2003) 387}
  [\href{https://arxiv.org/abs/hep-ph/0308008}{{\ttfamily hep-ph/0308008}}].

\bibitem{AtwoodSoni:InPreparation}
D.~Atwood and A.~Soni, \emph{{unpublished}}, .

\bibitem{DAmbrosio:1998gur}
G.~D'Ambrosio, G.~Ecker, G.~Isidori and J.~Portoles, \emph{{The Decays K
  ---\ensuremath{>} pi l+ l- beyond leading order in the chiral expansion}},
  \href{https://doi.org/10.1088/1126-6708/1998/08/004}{\emph{JHEP} {\bfseries
  08} (1998) 004} [\href{https://arxiv.org/abs/hep-ph/9808289}{{\ttfamily
  hep-ph/9808289}}].

\bibitem{DAmbrosio:2018ytt}
G.~D'Ambrosio, D.~Greynat and M.~Knecht, \emph{{On the amplitudes for the
  CP-conserving $K^\pm(K_S)\to\pi^\pm(\pi^0)\ell^+\ell^-$ rare decay modes}},
  \href{https://doi.org/10.1007/JHEP02(2019)049}{\emph{JHEP} {\bfseries 02}
  (2019) 049} [\href{https://arxiv.org/abs/1812.00735}{{\ttfamily
  1812.00735}}].

\bibitem{Greenlee:1990qy}
H.B.~Greenlee, \emph{{Background to $K^0_L \to \pi^0 e e$ From $K^0_L \to
  \gamma \gamma e e$}},
  \href{https://doi.org/10.1103/PhysRevD.42.3724}{\emph{Phys. Rev. D}
  {\bfseries 42} (1990) 3724}.

\bibitem{ParticleDataGroup:2022pth}
{\scshape Particle Data Group} collaboration, \emph{{Review of Particle
  Physics}}, \href{https://doi.org/10.1093/ptep/ptac097}{\emph{PTEP} {\bfseries
  2022} (2022) 083C01}.

\bibitem{AlvesJunior:2018ldo}
A.A.~Alves~Junior et~al., \emph{{Prospects for Measurements with Strange
  Hadrons at LHCb}}, \href{https://doi.org/10.1007/JHEP05(2019)048}{\emph{JHEP}
  {\bfseries 05} (2019) 048}
  [\href{https://arxiv.org/abs/1808.03477}{{\ttfamily 1808.03477}}].

\bibitem{HIKE:2022qra}
{\scshape HIKE} collaboration, \emph{{HIKE, High Intensity Kaon Experiments at
  the CERN SPS}: {Letter of Intent}},
  \href{https://arxiv.org/abs/2211.16586}{{\ttfamily 2211.16586}}.

\bibitem{HIKE:2023ext}
{\scshape HIKE} collaboration, \emph{{High Intensity Kaon Experiments (HIKE) at
  the CERN SPS Proposal for Phases 1 and 2}},
  \href{https://arxiv.org/abs/2311.08231}{{\ttfamily 2311.08231}}.

\bibitem{Isidori:2004rb}
G.~Isidori, C.~Smith and R.~Unterdorfer, \emph{{The Rare decay $K_L\to \pi^0
  \mu^+ \mu^-$ within the SM}},
  \href{https://doi.org/10.1140/epjc/s2004-01879-0}{\emph{Eur. Phys. J. C}
  {\bfseries 36} (2004) 57}
  [\href{https://arxiv.org/abs/hep-ph/0404127}{{\ttfamily hep-ph/0404127}}].

\bibitem{DAmbrosio:2022kvb}
G.~D'Ambrosio, A.M.~Iyer, F.~Mahmoudi and S.~Neshatpour, \emph{{Anatomy of kaon
  decays and prospects for lepton flavour universality violation}},
  \href{https://doi.org/10.1007/JHEP09(2022)148}{\emph{JHEP} {\bfseries 09}
  (2022) 148} [\href{https://arxiv.org/abs/2206.14748}{{\ttfamily
  2206.14748}}].

\bibitem{SchachtSoni:InPreparation}
S.~Schacht and A.~Soni, \emph{{in preparation}}, .

\bibitem{Christ:2015aha}
{\scshape RBC, UKQCD} collaboration, \emph{{Prospects for a lattice computation
  of rare kaon decay amplitudes: $K\to\pi\ell^+\ell^-$ decays}},
  \href{https://doi.org/10.1103/PhysRevD.92.094512}{\emph{Phys. Rev. D}
  {\bfseries 92} (2015) 094512}
  [\href{https://arxiv.org/abs/1507.03094}{{\ttfamily 1507.03094}}].

\bibitem{Heiliger:1992uh}
P.~Heiliger and L.M.~Sehgal, \emph{{Analysis of the decay K-L ---\ensuremath{>}
  pi0 gamma gamma and expectations for the decays K(l) ---\ensuremath{>} pi0 e+
  e- and K(l) ---\ensuremath{>} pi0 mu+ mu-}},
  \href{https://doi.org/10.1103/PhysRevD.47.4920}{\emph{Phys. Rev. D}
  {\bfseries 47} (1993) 4920}.

\bibitem{Donoghue:1994yt}
J.F.~Donoghue and F.~Gabbiani, \emph{{Reanalysis of the decay K(L)
  ---\ensuremath{>} pi0 e+ e-}},
  \href{https://doi.org/10.1103/PhysRevD.51.2187}{\emph{Phys. Rev. D}
  {\bfseries 51} (1995) 2187}
  [\href{https://arxiv.org/abs/hep-ph/9408390}{{\ttfamily hep-ph/9408390}}].

\bibitem{Diwan:2001sg}
M.V.~Diwan, H.~Ma and T.L.~Trueman, \emph{{Muon decay asymmetries from K0(L)
  ---\ensuremath{>} pi0 mu+ mu- decays}},
  \href{https://doi.org/10.1103/PhysRevD.65.054020}{\emph{Phys. Rev. D}
  {\bfseries 65} (2002) 054020}
  [\href{https://arxiv.org/abs/hep-ph/0112350}{{\ttfamily hep-ph/0112350}}].

\bibitem{Mescia:2006jd}
F.~Mescia, C.~Smith and S.~Trine, \emph{{K(L) ---\ensuremath{>} pi0 e+ e- and
  K(L) ---\ensuremath{>} pi0 mu+ mu-: A Binary star on the stage of flavor
  physics}}, \href{https://doi.org/10.1088/1126-6708/2006/08/088}{\emph{JHEP}
  {\bfseries 08} (2006) 088}
  [\href{https://arxiv.org/abs/hep-ph/0606081}{{\ttfamily hep-ph/0606081}}].

\bibitem{Crivellin:2016vjc}
A.~Crivellin, G.~D'Ambrosio, M.~Hoferichter and L.C.~Tunstall, \emph{{Violation
  of lepton flavor and lepton flavor universality in rare kaon decays}},
  \href{https://doi.org/10.1103/PhysRevD.93.074038}{\emph{Phys. Rev. D}
  {\bfseries 93} (2016) 074038}
  [\href{https://arxiv.org/abs/1601.00970}{{\ttfamily 1601.00970}}].

\bibitem{DAmbrosio:2023irq}
G.~D'Ambrosio, F.~Mahmoudi and S.~Neshatpour, \emph{{Beyond the Standard Model
  prospects for kaon physics at future experiments}},
  \href{https://arxiv.org/abs/2311.04878}{{\ttfamily 2311.04878}}.

\bibitem{DAmbrosio:2017klp}
G.~D'Ambrosio and T.~Kitahara, \emph{{Direct $CP$ Violation in $K \to \mu^+
  \mu^-$}}, \href{https://doi.org/10.1103/PhysRevLett.119.201802}{\emph{Phys.
  Rev. Lett.} {\bfseries 119} (2017) 201802}
  [\href{https://arxiv.org/abs/1707.06999}{{\ttfamily 1707.06999}}].

\bibitem{Dery:2021mct}
A.~Dery, M.~Ghosh, Y.~Grossman and S.~Schacht, \emph{{$K \to {\mu}^{+} \mu^-$
  as a clean probe of short-distance physics}},
  \href{https://doi.org/10.1007/JHEP07(2021)103}{\emph{JHEP} {\bfseries 07}
  (2021) 103} [\href{https://arxiv.org/abs/2104.06427}{{\ttfamily
  2104.06427}}].

\bibitem{Brod:2022khx}
J.~Brod and E.~Stamou, \emph{{Impact of indirect CP violation on
  Br(K$_{S}$\textrightarrow{}
  \ensuremath{\mu}$^{+}$\ensuremath{\mu}$^{-}$)$_{l=0}$}},
  \href{https://doi.org/10.1007/JHEP05(2023)155}{\emph{JHEP} {\bfseries 05}
  (2023) 155} [\href{https://arxiv.org/abs/2209.07445}{{\ttfamily
  2209.07445}}].

\bibitem{Dery:2022yqc}
A.~Dery, M.~Ghosh, Y.~Grossman, T.~Kitahara and S.~Schacht, \emph{{A Precision
  Relation between $\Gamma(K\to\mu^+\mu^-)(t)$ and ${\cal
  B}(K_L\to\mu^+\mu^-)/{\cal B}(K_L\to\gamma\gamma)$}},
  \href{https://doi.org/10.1007/JHEP03(2023)014}{\emph{JHEP} {\bfseries 03}
  (2023) 014} [\href{https://arxiv.org/abs/2211.03804}{{\ttfamily
  2211.03804}}].

\bibitem{Schacht:2023yav}
S.~Schacht, \emph{{Kaon Decays beyond the Standard Model}},  in \emph{{57th
  Rencontres de Moriond on QCD and High Energy Interactions}}, 5, 2023
  [\href{https://arxiv.org/abs/2305.06267}{{\ttfamily 2305.06267}}].

\bibitem{Marchevski:2023kab}
R.~Marchevski, \emph{{First thought on a high-intensity KS experiment}},
  \href{https://doi.org/10.1088/1742-6596/2446/1/012035}{\emph{J. Phys. Conf.
  Ser.} {\bfseries 2446} (2023) 012035}
  [\href{https://arxiv.org/abs/2301.06801}{{\ttfamily 2301.06801}}].

\bibitem{LHCb:2020ycd}
{\scshape LHCb} collaboration, \emph{{Constraints on the $K^0_S \rightarrow
  \mu^+ \mu^-$ Branching Fraction}},
  \href{https://doi.org/10.1103/PhysRevLett.125.231801}{\emph{Phys. Rev. Lett.}
  {\bfseries 125} (2020) 231801}
  [\href{https://arxiv.org/abs/2001.10354}{{\ttfamily 2001.10354}}].

\bibitem{Dery:2021vql}
A.~Dery and M.~Ghosh, \emph{{K \textrightarrow{}
  \ensuremath{\mu}$^{+}$\ensuremath{\mu}$^{-}$ beyond the standard model}},
  \href{https://doi.org/10.1007/JHEP03(2022)048}{\emph{JHEP} {\bfseries 03}
  (2022) 048} [\href{https://arxiv.org/abs/2112.05801}{{\ttfamily
  2112.05801}}].

\bibitem{Christ:2022rho}
N.~Christ, X.~Feng, L.~Jin, C.~Tu and Y.~Zhao, \emph{{Lattice QCD Calculation
  of \ensuremath{\pi}0\textrightarrow{}e+e- Decay}},
  \href{https://doi.org/10.1103/PhysRevLett.130.191901}{\emph{Phys. Rev. Lett.}
  {\bfseries 130} (2023) 191901}
  [\href{https://arxiv.org/abs/2208.03834}{{\ttfamily 2208.03834}}].

\bibitem{Christ:2020bzb}
N.H.~Christ, X.~Feng, L.~Jin, C.~Tu and Y.~Zhao, \emph{{Lattice QCD calculation
  of the two-photon contributions to $K_L \to \mu^+ \mu^-$ and $\pi^0 \to e^+
  e^-$ decays}}, \href{https://doi.org/10.22323/1.363.0128}{\emph{PoS}
  {\bfseries LATTICE2019} (2020) 128}.

\bibitem{Buras:2021nns}
A.J.~Buras and E.~Venturini, \emph{{Searching for New Physics in Rare $K$ and
  $B$ Decays without $|V_{cb}|$ and $|V_{ub}|$ Uncertainties}},
  \href{https://doi.org/10.5506/APhysPolB.53.6-A1}{\emph{Acta Phys. Polon. B}
  {\bfseries 53} (2021) 6} [\href{https://arxiv.org/abs/2109.11032}{{\ttfamily
  2109.11032}}].

\bibitem{Grossman:1997sk}
Y.~Grossman and Y.~Nir, \emph{{K(L) ---\ensuremath{>} pi0 neutrino
  anti-neutrino beyond the standard model}},
  \href{https://doi.org/10.1016/S0370-2693(97)00210-4}{\emph{Phys. Lett. B}
  {\bfseries 398} (1997) 163}
  [\href{https://arxiv.org/abs/hep-ph/9701313}{{\ttfamily hep-ph/9701313}}].

\end{thebibliography}\endgroup

\end{document}